\documentclass[aps,prx,twocolumn,superscriptaddress,nofootinbib,notitlepage]{revtex4-1}
\usepackage{amsmath}
\usepackage{amstext}
\usepackage{amssymb}
\usepackage{xfrac}
\usepackage{times}
\usepackage[colorlinks,citecolor=blue]{hyperref}
\usepackage{graphicx}
\usepackage{amsmath}
\usepackage{amstext}
\usepackage{amssymb}
\usepackage{amsfonts}
\usepackage{longtable,booktabs}
\usepackage{hyperref}\usepackage{url}
\usepackage{subfigure}%
\usepackage{dsfont}
\usepackage{booktabs}
\usepackage{amsbsy}
\usepackage{dcolumn}
\usepackage{amsthm}
\usepackage{bm}
\usepackage{esint}
\usepackage{multirow}
\usepackage{hyperref}
\hypersetup{
    colorlinks=true,
    linkcolor=blue,
    filecolor=magenta,
    urlcolor=blue,
}
\usepackage{cleveref}

\usepackage{mathrsfs}
\usepackage{amsfonts}
\usepackage{amsbsy}
\usepackage{dcolumn}
\usepackage{bm}
\usepackage{multirow}
\usepackage{color}

  \usepackage{extarrows}
\usepackage{datetime}

\usepackage{comment}
\usepackage[super]{nth}

\makeatother

\begin{document}
\title{Hierarchical Proliferation of Higher-Rank Symmetry Defects in  Fractonic Superfluids}
\author{Jian-Keng Yuan}
\affiliation{School of Physics, State Key Laboratory of Optoelectronic Materials
and Technologies, and Guangdong Provincial Key Laboratory of Magnetoelectric
Physics and Devices, Sun Yat-sen University, Guangzhou, 510275, China}
\author{Shuai A. Chen}
\email{chsh@ust.hk}

\affiliation{Department of Physics, The Hong Kong University of Science and Technology,
Hong Kong SAR, China}
\author{Peng Ye}
\email{yepeng5@mail.sysu.edu.cn}

\affiliation{School of Physics, State Key Laboratory of Optoelectronic Materials
and Technologies, and Guangdong Provincial Key Laboratory of Magnetoelectric
Physics and Devices, Sun Yat-sen University, Guangzhou, 510275, China}

\begin{abstract}
Symmetry defects, e.g., vortices in conventional superfluids, play a critical role in a complete description of symmetry-breaking phases. In this paper, we develop the theory of symmetry defects in fractonic superfluids, i.e., spontaneously higher-rank
symmetry (HRS) breaking phases.  
By Noether's theorem, HRS is associated with the conservation law of higher moments, e.g. dipoles, quadrupoles, and angular moments. 
We   establish   finite-temperature phase diagrams by identifying a series
of topological phase transitions via the renormalization group  
flow equations and Debye-H\"uckel approximation. Accordingly, a series of Kosterlitz-Thouless  topological
transitions are found to occur successively at different temperatures, which are triggered by  proliferation of  defects, defect bound states, and so on. Such a \emph{hierarchical proliferation} brings rich phase structures.  
Meanwhile, a screening effect from sufficiently high density of defect bound states leads to instability and collapse of the intermediate temperature phases, which further enriches the phase diagrams. For concreteness, we consider a fractonic superfluid  in which      ``angular moments'' are conserved. We then present the general theory, in which other types of HRS   can be analyzed in a similar manner.   
Further   directions are present at the end of the paper. 
\end{abstract}
\date{\today}

\maketitle


\section{Introduction}

The concept of symmetry and its spontaneous breaking has influenced quantum physics, from phases and phase transitions to the Standard
Model of fundamental forces. A symmetry breaking phase can be characterized
by an order parameter that varies with respect to symmetry operation and fluctuates in the vicinity of mean-field configurations. Distinct from smooth fluctuations, such
as phonons and rotons, a singularity  can be induced  by symmetry 
defects, which tends to disorder the phase and restore symmetry. In a conventional 2D 
superfluid where U(1) global symmetry related to particle number conservation gets spontaneously broken, for example, the symmetry defects ``superfluid vortex'' at sufficiently low temperatures   are confined into   dipole-like bound states
such that the superfluid phase is sustained against thermal fluctuations. At a critical value $T_{c}$, vortices begin to be released from bound states, which eventually results in a vortex proliferation  that    destroys the superfluid
phase and restores symmetry at temperatures higher than $T_c$. This is one of great achievements that lead to 2016 Nobel Prize in Physics, namely, celebrated Kosterlitz-Thouless
(KT) physics~\cite{Kosterlitz_1973,Kosterlitz2016,chaikin2000principles},
which provides one of most influential prototypes of topological phase
transitions driven by symmetry defects.

As symmetry is an indispensable ingredient in the above KT physics, the   goal of this paper is to explore KT-like physics in many-body systems where \emph{higher-rank symmetry} (HRS) gets spontaneously broken.  Let us first give a brief introduction to HRS.  Electromagnetism textbook tells us, by starting with the particle density $\rho$, 
we can express a series of multipole moments, 
such as 
$\rho x^a, \rho x^a x^b,\cdots$. While  the conservation of charge, i.e., the integration of $\rho$ over the whole space, is a very familiar fact in, e.g., metals and insulators,  it is also meaningful for considering systems with conserved higher moments. Furthermore, Noether's theorem   indicates that each of conserved quantities here must be associated with an underlying continuous symmetry, which is  nothing but  aforementioned HRS.  In the literature, the condensed matter physics of such type of symmetry has  been initiated and paid great attentions to in the field of   fracton physics~\cite{Nandkishore2019,2020Fracton} where diverse quantum phenomena have been discovered, including system-size dependent noise-immune ground state
degeneracy, exotic quantum dynamics, see, e.g., Refs.~\cite{Chamon05,Haah2011,Vijay2015,Vijay2016,Pretko2017,Ma2017,PhysRevB.103.155140,2019arXiv191014048M,Shirley2018,PhysRevResearch.2.033021,ye19a,PhysRevB.104.235127,QMC_Zhou,Zhou:2022eig,Li:2022xwk}.  
As a side, while there are   alternative terminologies in the literature, here we call these symmetries ``higher rank symmetry'' due to the resultant 
higher-rank tensor gauge field after gauging it.  

\begin{figure}
\centering \includegraphics[scale=0.5]{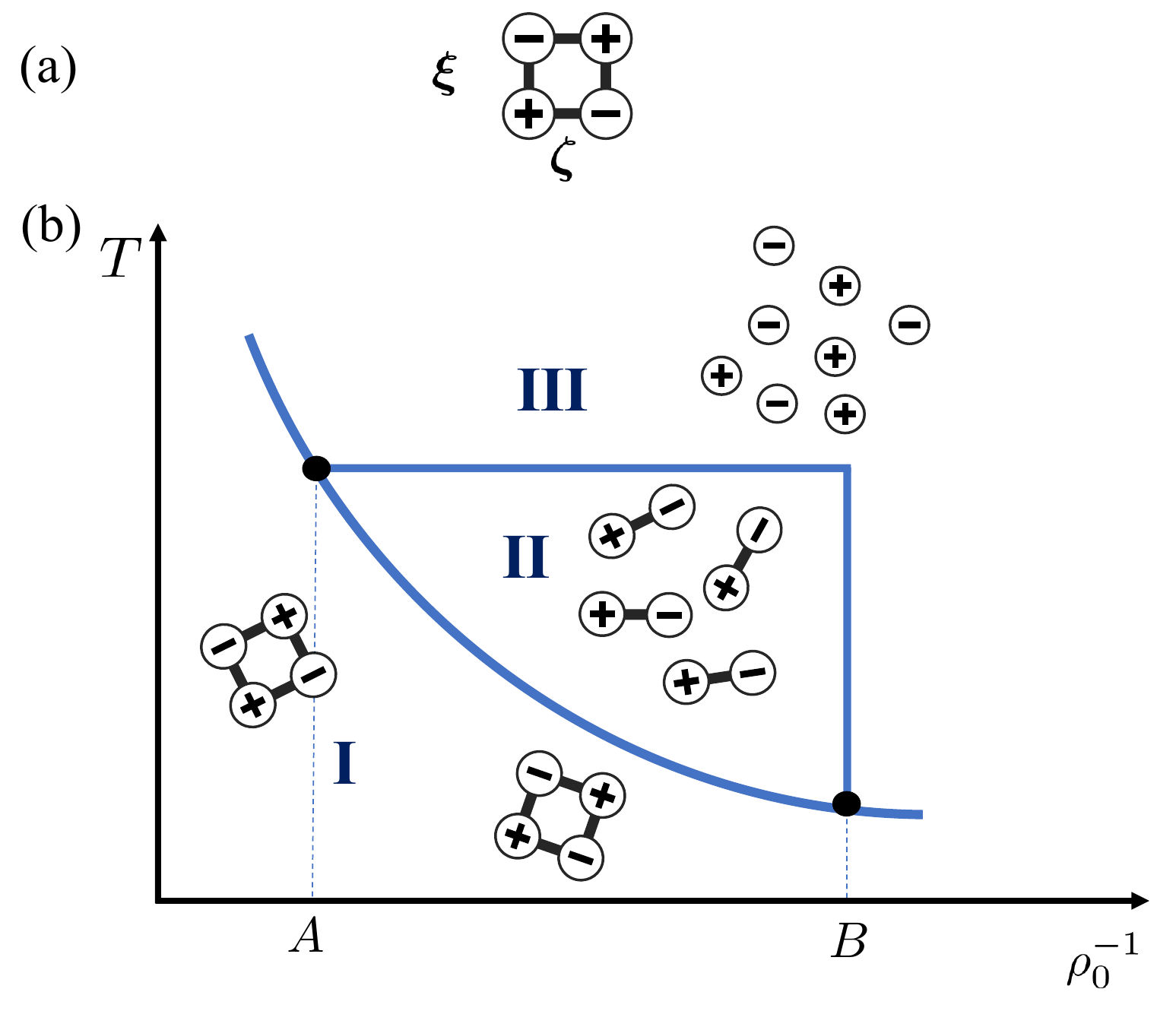} 
\caption{Global phase diagram.  (a) A quadrupole bound state  of a pair of a unit defect and a unit anti-defect. 
 (b) Finite-temperature phase diagram from the renormalization group analysis and Debye-H\"uckel approximation. $T$ and $\rho_0^{-1}$ are respectively  temperature and superfluid stiffness inverse.  In Phase-I, defects $\Theta$ are confined in the form of quadrupole
bound states. In Phase-II, defects are confined in the form of dipole bound states. In Phase-III, defects are fully released from bound states and are wildly mobile. The locations of $A$ and $B$ depend on core energy values of both  defects and dipole bound states (see the main text). Two  topological  transitions occur successively from Phase-I, Phase-II, to Phase-III. A direct transition from Phase-I to Phase-III occurs if $\rho^{-1}_0$ is outside the domain $(A, B)$.}
\label{Fig:phasediagram} 
\end{figure} 
By noting that there have been great triumphs in the conventional spontaneously symmetry-breaking phase, we are motivated to ask if there  is  novel   many-body physics     by instead considering HRS. For example, in the framework of HRS, what can we expect for the superfluid-like phase, off-diagonal long-range order, Mermin-Wagner-like theorem, symmetry defects, and KT-like physics? Along this line of thinking,  Refs.~\cite{2020PhRvR2b3267Y,Chen2021FS,Ye2021PRR,Quantum2022Yuan}
construct concrete minimal models for realizing unconventional
ordered phases dubbed \textit{fractonic superfluids}, in
which higher-rank $U(1)$ symmetry gets spontaneously broken in various
fashions. Many exotic phenomena beyond the conventional symmetry-breaking phases have been identified in various channels~\cite{2020PhRvR2b3267Y,Chen2021FS,Ye2021PRR,Quantum2022Yuan,2022PhRvB.105o5107S,2022arXiv220809056K,PhysRevD.104.105001}. In this paper, we will focus on   HRS defects and HRS-generalization of KT physics in fractonic superfluids.

In this paper, we   establish a general framework of finite-temperature phase diagrams
and a series of KT topological phase transitions driven by proliferation of either HRS symmetry defects or defect bound states.  We discover that, thermally activated  HRS defects generally constitute a hierarchy of bound states, such as 
dipole with neutral charge [{\textbf{Terminology clarification}:  \textit{Hereafter, unless otherwise specified, charge means the vorticity/winding number of HRS defect, while dipole, quadrupole, etc. really are formed by defects rather than original condensed bosons.}}], quadrupole [Fig.~\ref{Fig:phasediagram}(a)] with both neutral charge and
dipole moment, and so on. They bear a series of self-energy divergence,
e.g., $\ln L$,$L$,$L^{2},\cdots$, with $ L$ being the system
size. Upon increasing temperature, various bound states successively get dissolved, from which new bound states of defects are successively released and then proliferate. After a series of   KT transitions, HRS defects are finally released from all  bound states and become mobile
  at   high temperatures. 

 We take the conservation of so-called \emph{angular   moments} as a concrete example~\cite{Chen2021FS,Ye2021PRR} and then go to general theory. Through renormalization-group (RG) analysis and  Debye-H\"uckel approximation, we identify
two characteristic KT transitions and three distinct phases as shown in Fig.~\ref{Fig:phasediagram}(b).  
In   Phase-I,
defects are confined into quadrupole bound states such that superfluidity is sustained. 
If further increasing temperature with $\rho^{-1}_0$ being in the domain $(A,B)$, Phase-I is transited into Phase-II at $T_{c1}$, which is the critical temperature of the first KT transition where quadrupole bound states get dissolved and  dipole bound states get proliferated. In Phase-II,  superfluidity is destroyed. Then, if temperature is further increased, dipole bound states form a plasma with high density which drastically renormalizes the bare interaction between defects. As a result, the second KT transition occurs at $T_{c2}$, which sends the system into Phase-III.  As the defects are no more bound in the form of BS, defects are wildly mobile in Phase-III, which is similar to the episode in the movie \emph{Pirates of the Caribbean}: the enraged goddess Calypso, once being released from human form, generates a massive maelstrom.    On the other hand, if $\rho^{-1}_0$ is either too small or too large, Phase-I is directly transited to Phase-III in Fig.~\ref{Fig:phasediagram}.

The remainder of this paper is organized as follows. In Sec.~\ref{section_model}, we concretely set up the minimal model, higher-rank symmetry, and defect configurations. In Sec.~\ref{section_phase1}, we study Phase-I where defects are confined in quadrupole-like bound states. In Sec.~\ref{section_phase2}, we study Phase-II where defects are confined in dipole-like bound states. Key results are summarized in Sec.~\ref{section_phase2_suppl}. In Sec.~\ref{section_phase3}, we study Phase-III where defects are deconfined. Key results are summarized in Sec.~\ref{section_phase3_suppl}. In Sec.~\ref{section_generalization}, we generalize the above study to more complex higher-rank symmetry. In The paper is concluded in Sec.~\ref{section_summary}.  

\section{Minimal model, higher-rank symmetry, and defect configurations}\label{section_model}

We start with the following minimal Hamiltonian  that conserves angular moments~\cite{Chen2021FS,Ye2021PRR,Quantum2022Yuan},
\begin{equation}
 {H}=\vert\partial_{1}\hat{\Phi}_{1}\vert^{2}+\vert\partial_{2}\hat{\Phi}_{2}\vert^{2}+\hat{V}_{0}+\hat{V}_{1}~,
 \label{eq:Ham}
\end{equation}
where 
\begin{align}
\hat{V}_{0}=\sum_{\alpha=1}^{2}-\mu\hat{\rho}_{\alpha}+\frac{g}{2}\hat{\rho}_{\alpha}^{2}
\end{align}
 and 
 \begin{align}
 \hat{\rho}_{\alpha}=\hat{\Phi}_{\alpha}^{\dagger}\hat{\Phi}_{\alpha}\,.
 \end{align} 
Here $\hat\Phi_{1,2}$ denote two components of complex bosonic fields. $\hat{V}_{0}$ includes a  repulsive interaction and a chemical potential term. 
The term 
\begin{align}\hat{V}_{1}=K\vert\hat{\Phi}_{1}\partial_{1}\hat{\Phi}_{2}+\hat{\Phi}_{2}\partial_{2}\hat{\Phi}_{1}\vert^{2}
\end{align}
represents a two-particle correlated hopping process.
Apart from the particle number conservation for each component, the
model  has an extra $U(1)$ to conserve the total \emph{angular
moments} that are a vectorial generalization of dipoles: 
\begin{equation}
Q_{12}=\int d^{2}\mathbf{r}(\hat{\rho}_{1}y-\hat{\rho}_{2}x)\,,
\end{equation}
where $\mathbf{r}=(x,y)$.  Indeed, one may check that
\begin{align}
[Q_{12},   {H}]=0
\end{align} upon applying canonical quantization and regarding $Q_{12}$ as a generator of HRS. Thus, a group element of HRS can be expressed as $e^{i\theta Q_{12}}$. 
Here we discard    symmetry-respecting terms with more complicated momentum dependence as all these terms are irrelevant at low energies.
Once $\hat{V}_0$ forms a Mexican hat, we obtain a \emph{fractonic superfluid} phase with off-diagonal long-range order  and non-vanishing
order parameters 
 $\langle\hat{\Phi}_{1}\rangle=\langle\hat{\Phi}_{2}\rangle\neq0 
 $  
whose low-energy  effective Hamiltonian reads  
\begin{align}
\!{H}[\theta_1,\theta_2]=\frac{1}{2}\rho_{0}[(\partial_{1}\theta_{1})^{2}+(\partial_{2}\theta_{2})^{2}+(\partial_{1}\theta_{2}+\partial_{2}\theta_{1})^{2}],\label{eq:Goldstone mode}
\end{align}
where $\theta_{1,2}$ are Goldstone modes. Without loss of generality, we set $K\rho_{0}=1$ with $\rho_{0}$ being  the superfluid stiffness, which simplifies the formulas. 
As
the velocity fields $v_{\alpha\beta}$\cite{Quantum2022Yuan} ($\alpha,\beta=1,2$)
of $\alpha$th component bosons of $\beta$th direction in fractonic
superfluids are
\begin{equation}
v_{\alpha\beta}=\delta_{\alpha\beta}\partial_{\alpha}\theta_{\alpha}+(1-\delta_{\alpha\beta})(\partial_{\alpha}\theta_{\beta}+\partial_{\beta}\theta_{\alpha}),\label{velocity}
\end{equation}
the Hamiltonian in Eq.~(\ref{eq:Goldstone mode}) reduces to the
following compact form 
\begin{equation}
H=\frac{\rho_{0}}{2}\int d^{2}\mathbf{r}\sum_{\alpha\beta}v_{\alpha\beta}^{2}.\label{Hamiltonian: velocity}
\end{equation}

The bare superfluid stiffness $\rho_{0}$, a hallmark of the superfluid
phase, can be reduced by thermal defects. In general, due to thermally
activated topological defects, we can decompose the fields $\theta_{\alpha}$
into a smooth part $\tilde{\theta}_{\alpha}$ and singular part $\theta_{\alpha}^{s}$,
i.e. 
 $\theta_{\alpha}=\tilde{\theta}_{\alpha}+\theta_{\alpha}^{s}.
 $ 
And the velocity field follows a similar decomposition, the smooth
part $\tilde{v}_{\alpha\beta}$ describing the superfluidity and the
singularity part $v_{s,\alpha\beta}$ describing topological defects,
\begin{equation}
v_{\alpha\beta}=\tilde{v}_{\alpha\beta}+v_{s,\alpha\beta},\label{decomposition: velocity}
\end{equation}
where both of the components $\tilde{v}_{\alpha\beta}$ and
$v_{s,\alpha\beta}$ obey the expression of velocity fields in Eq.~(\ref{velocity}).
Symmetry defects denoted as $\Theta$ can arise due to the compactness of $\theta_{1,2}$, which have the following general solutions: 
\begin{equation}
\!\!\!\!\!\Theta\equiv(\theta_{1,}\theta_{2})\!=\!\ell\left(\frac{y}{a}\varphi(\mathbf{r})-\frac{x}{a}\ln\frac{r}{a},-\frac{x}{a}\varphi(\mathbf{r})-\frac{y}{a}\ln\frac{r}{a}\right).\!\!\!\label{rank-1}
\end{equation}
Here $a$ is the size of the defect core and a multi-valued function
$\varphi(\mathbf{r})$ is the angle of $\mathbf{r}$ relative
to the core. The defect charge $\ell$ is quantized as a momentum that is well-understood on a lattice. The general   rules of constructing symmetry defects have been established in Ref.~\cite{Chen2021FS}. 




 \section{Phase-I: defects are confined in  quadrupoles}\label{section_phase1}
We consider $N$ defects $\{\Theta_i\}$  carrying winding number (i.e., ``charge'') $\ell_{i}$
with the core coordinates $\mathbf{r}_{i}$ ($i=1,\ldots,N$), and
then the total energy can be expressed in the momentum space:
\begin{align}
\mathscr{E}[\{\Theta_i\}]=\frac{\rho_{0}}{2}\sum_{i,j=1}^{N}\ell_{i}\ell_{j}\int d^{2}\mathbf{q}\mathscr{U}_\mathbf{q}e^{i\mathbf{q}\cdot(\mathbf{r}_{i}-\mathbf{r}_{j})}\,,
\end{align}
where $\mathscr{U}_\mathbf{q}=\frac{2}{q^{4}}$ and the momentum $q=|\mathbf{q}|$. 
  In this expression, a series of severely infrared divergent terms can be identified. By observing
\begin{equation}
\!\!\!\!\int \!d^{2}\mathbf{q}\mathscr{U}_\mathbf{q}e^{i\mathbf{q}\cdot(\mathbf{r}_{i}-\mathbf{r}_{j})}\!=\!2\pi\bigg(\frac{L}{a}\bigg)^{2}\!- \pi\frac{\vert\mathbf{r}_{i}-\mathbf{r}_{j}\vert^{2}}{a^{2}}\ln\!\frac{L}{a}+\! f,\label{eq:U11}
\end{equation}
we can single out two types of divergence, i.e., $L^{2}$ and $\ln L$,
while the letter $f$ incorporates all finite terms. In the low temperature
region, we have to impose the constraints on the defect configuration
for energetic consideration. Suppression of the $L^{2}$ divergence requires   the net charges vanish, i.e., 
\begin{align}\sum_{i=1}^{N}\ell_{i}=0\,.
\end{align}
Under the charge neutral condition,
  suppression of $\ln L$ divergence further requires    the net dipole moments vanish, i.e., 
\begin{align}\sum_{i=1}^{N}\ell_{i}\mathbf{r}_{i}  =0\,.
\end{align}
Once both conditions 
are satisfied, all divergence
in Eq.~\eqref{eq:U11} disappears, yielding the screened interacting energy: 
\begin{equation}
\mathcal{I}[\Theta_i,\Theta_j]=\pi\rho_{0}\sum_{i,j}^{N}\ell_{i}\ell_{j}\frac{\vert\mathbf{r}_{i}-\mathbf{r}_{j}\vert^{2}}{a^{2}}\ln\frac{\vert\mathbf{r}_{i}-\mathbf{r}_{j}\vert}{a}~.\label{eq:screenedU11}
\end{equation}
Despite being canceled out, the divergence in  Eq.~(\ref{eq:U11}) physically allows us to respectively define  self-energy formulas of three distinct objects, namely, a single defect, a dipole bound state (BS), and a quadrupole BS: 
\begin{enumerate}
\item \textbf{Divergence of self-energy of a single defect.} The self-energy ($\mathscr{E}^s_{\mathrm{self}}$) of a single defect is polynomially divergent ($\sim L^2$) as $L\rightarrow \infty$: 
\begin{equation}
\mathscr{E}^s_{\mathrm{self}}=\pi\rho_{0}\ell^{2}\left(\frac{L}{a}\right)^{2}\longrightarrow \infty\, 
\end{equation}
which is apparently different from the logarithmic divergent self-energy of vortices (i.e., defects) in the conventional KT physics.

\item \textbf{Divergence of self-energy of a  dipole bound state.} The self-energy 
($\mathscr{E}^d_{\mathrm{self}}$) of  a dipole BS formed by 
two defects with opposite charges $\pm\ell$ is logarithmically divergent ($\sim \ln L$) as $L\rightarrow \infty$:
\begin{equation}
\mathscr{E}^d_{\mathrm{self}}=\pi\rho_{0}\left(\frac{p}{a}\right)^{2}\ln\frac{L}{a}\longrightarrow \infty
\label{eq:dbs_self}
\end{equation} {}
with $p=|\mathbf{p}|$ and $\mathbf{p}=\ell\boldsymbol{\xi}\equiv\ell(\mathbf{r}_{1}-\mathbf{r}_{2})$ being
the dipole moment. Thus, from the expression of self-energy, a dipole BS here resembles a vortex  in the conventional KT physics.

\item \textbf{Finiteness of self-energy of a  quadrupole bound state.} A  quadrupole BS is formed by four defects with vanishing  charges and vanishing dipole moments, as shown in Fig.~\ref{Fig:phasediagram}(a). Its   self-energy ($\mathscr{E}^q_{\mathrm{self}}$) is given by ($\Delta_{\pm}\equiv \vert\boldsymbol{\xi}\pm\boldsymbol{\zeta}\vert$, $\xi=|\boldsymbol{\xi}|$, $\zeta=|\boldsymbol{\zeta}|$):
\begin{align}
\mathscr{E}^q_{\mathrm{self}}\!= \! \pi\rho_{0}\bigg(\! \sum_{\iota=\pm}\!\frac{\Delta_\iota^{2}}{a^{2}}\!\ln\frac{\Delta_\iota}{a}\!  -\!2\!\!\sum_{\tau=\xi,\zeta}\frac{\tau^{2}}{a^{2}}\!\ln\frac{\tau}{a}\bigg)\!\!<\!\!\infty.\!\!\!\!\!\!\!\!\!\!\!
\end{align}
The finiteness of this self-energy  is expected since  the definition of  quadrupole BSs exactly meets  the aforementioned two conditions.
\end{enumerate}
Bearing the above analysis in mind, we conclude that, at the thermodynamic limit, HRS defects  are energetically
confined into quadrupole BSs at low temperatures (Phase-I in Fig.~\ref{Fig:phasediagram}). Phase-I shows the algebraic long-range order since the correlation function $\langle \hat\Phi(\mathbf r_1)\hat\Phi^\dagger(\mathbf r_2)\rangle$ behaves as a power-law function at long distances due to the   gapless phonon mode excitations with a   stiffness renormalized by defects.

 \section{Phase-II: defects are confined in   dipoles} \label{section_phase2}
\subsection{General properties of Phase-II}\label{section_phase2_suppl}
When the temperature is increased to some critical value $T_{c1}$, dipole BSs are completely released from quadrupole BSs and proliferate wildly, which leads to a topological phase transition (named as ``the first KT transition'') from Phase-I to Phase-II.  In Phase-II, the superfludity is fully destroyed, and thus we do not need to consider the gapless phonon excitations. Instead, 
the system consists of mobile dipole BSs, while a few defects $\Theta$ can be thermally activated. 
Since a dipole BS is by definition formed by two $\Theta$'s with opposite charges, we can deduce the Hamiltonian of the interacting dipole BSs directly from Eq.~\eqref{eq:screenedU11}: 
\begin{equation}
H_{\text{D}}=\int d^{2}\mathbf{q}\sum_{\alpha\beta}\bigg(\frac{\rho_{0}q_{\alpha}q_{\beta}}{q^{4}}+\delta_{\alpha\beta}\mathcal{E}_{d}\bigg)D_{\alpha}(\mathbf{q})D_{\beta}(-\mathbf{q})~,\label{eq:dbsHam}
\end{equation}
where $\mathbf D(\mathbf q)\equiv (D_1(\mathbf q),D_2(\mathbf q))$ is the dipole BS density field.  $\mathcal{E}_{d}$ is the dipole BS core energy.  We leave the concrete derivation of this Hamiltonian to Sec.~\ref{section_HD}. Physically,  the first term in Eq.~(\ref{eq:dbsHam}) accounts for the self-energy divergence shown in Eq.~\eqref{eq:dbs_self} while the second term is the core energy. 
When temperature increases, dipole BSs reach high density, which will substantially renormalize   the interaction (\ref{eq:screenedU11}) between two bare defects $\Theta$.


To analytically understand the phase transition, 
we introduce the dipole BS fugacity $y_{d}\equiv\exp(-\beta\mathcal{E}_{d})$ where $\beta$ is the inverse temperature. In general, by definition, a
dipole BS is a stable BS object when $\mathcal{E}_{d}<2\mathcal{E}_{s}$ 
with $\mathcal{E}_{s}$ being the core energy of a single defect $\Theta$. The thermally activated 
dipole BSs can weaken the superfluid stiffness $\rho$, which can be justified via  the perturbative RG analysis. 
It should be noted that, in the RG calculation of the first KT transition,  it is enough to  keep track of   the renormalization effect of released dipole BSs on superfluidity since
an isolated defect $\Theta$ suffers from much severe energy divergence than a dipole BS. The $\beta$-equations for the superfluid stiffness $\rho$ and the fugacity $y_d$ are given by: 
\begin{align}
\frac{d\rho^{-1}}{dl} & =2\beta y_{d}^{2}+\mathcal{O}(y_{d}^{4})~,\label{eq:rhorg}\\
\frac{dy_{d}}{dl }& =(2-\pi\beta\rho)y_{d}+\mathcal{O}(y_{d}^{2})~,\label{eq:yeq}
\end{align}
where the dimensionless parameter $l$ parametrizes the RG flows. 
 We leave the detailed derivation of these two $\beta$-functions to Sec.~\ref{section_derivation_of_beta_functions}. 
As the right-hand side of Eq.~\eqref{eq:rhorg} is positive-definite, we can
find the critical temperature
\begin{equation}
T_{c1}=\frac{\pi\rho_0}{2} 
\end{equation} 
at the sign change point of the factor ``$(2-\pi\beta\rho)$'' in Eq.~(\ref{eq:yeq}).  Alternatively,  
$T_{c1}$ can be simply obtained  by a free energy argument in which sign change of the free-energy of a test object (i.e., an externally inserted dipole BS) occurs at $T_{c1}$. The derivation in the argument   is almost the same as that in the conventional KT physics except that the  {test object} here is not a vortex (defect) but a dipole BS.



\subsection{   Hamiltonian formalisms  in terms of  defect density fields\label{section_HD}}


The defects $\Theta$ are the building blocks for dipole and quadrupole
BSs. Hence, we can unify the description of defect interaction. Given a defect system, in general,
it contains isolated defects $\Theta$, dipole BSs and the
quadrupole BSs. For the sake of clarification, we particularly refer to the isolated defects $\Theta$ to distinguish ones that are confined in BSs. Nevertheless, a density field $s(\mathbf r)$ can be defined for the total defects $\Theta$, including both isolated and confined ones,
 $s(\mathbf{r})=\sum_{i}\ell_{i}\delta(\mathbf{r}-\mathbf{r}_{i})\,,
 $ 
where $\ell_{i}$ is the defect charge and $\mathbf{r}_{i}$ is the
position of the defect core. In this subsection, we will derive a unified Hamiltonian for defect interaction by means of the defect density $s(\mathbf r)$.

A dipole BS is formed by two neighboring defects $\Theta$ with opposite charges $\pm\ell$ that are separated by $\boldsymbol{\xi}$, 
\begin{align}
\ell\Theta(\mathbf{r})-\ell\Theta(\mathbf{r}-\boldsymbol{\xi}) & =\ell\boldsymbol{\xi}\cdot\nabla\Theta(\mathbf{r})\,.\label{eq:dipoleboundstate}
\end{align}
Here we use the notation $\Theta$ to denote a defect with charge $\ell =1$. 
Accordingly, we introduce the density of dipole BSs
 
\begin{equation}
\mathbf{D}(\mathbf{r})\equiv\sum_{i}\ell_{i}(\frac{\xi_{ix}}{a},\frac{\xi_{iy}}{a})\delta(\mathbf{r}-\mathbf{r}_{i})\equiv\sum_{i}\mathbf{p}_{i}\delta(\mathbf{r}-\mathbf{r}_{i})\,,
\end{equation}
while on a lattice, $\xi_{ix}/a$ and $\xi_{iy}/a$ take integer values.
 Here, $\mathbf{p}_{i}$ is the dipole moment of the $i$th dipole
BS.
From the relation in Eq.~(\ref{eq:dipoleboundstate}), we reach a condition 
 $s(\mathbf{r})=-a\nabla\cdot\mathbf{D}(\mathbf{r})\,,
 $ 
where $s(\mathbf r)$ represents the density of confined defects $\Theta$ in a dipole BS. 

Similarly, a quadrupole BS consists of four defects $\Theta$,
\begin{align}
 & \ell\Theta(\mathbf{r})-\ell\Theta(\mathbf{r}-\boldsymbol{\xi})-\ell\Theta(\mathbf{r}-\boldsymbol{\zeta})+\ell\Theta(\mathbf{r}-\boldsymbol{\xi}-\boldsymbol{\zeta})\nonumber \\
= & \sum_{\alpha,\beta}\ell\xi_{\alpha}\zeta_{\beta}\partial_{\alpha}\partial_{\beta}\Theta(\mathbf{r}),
\end{align}
where  $\alpha,\beta=1,2$ and
$\partial_{\alpha}\partial_{\beta}\Theta(\mathbf{r})$ represents
the configuration of a quadrupole BS. Thus, we can introduce
a quadrupole BS density, 
 ${Q}_{\alpha\beta}(\mathbf{r})=\sum_{i}\ell_{i}\frac{\xi_{i\alpha}\zeta_{i\beta}}{a^{2}}\delta(\mathbf{r}-\mathbf{r}_{i}),
 $  which can be related to the density of defect $\Theta$ via 
 $ s(\mathbf{r})=a^{2}\sum_{\alpha,\beta}\partial_{\alpha}\partial_{\beta}Q_{\alpha\beta}(\mathbf{r}).
 $

In summary, for a defect system, if we apply the notations $\bar{s}(\mathbf{r})$,
$\mathbf{\mathbf{D}(\mathbf{r})}$ and $ {Q}$ respectively
for isolated defect density, density of dipole BSs and density of
quadrupole BSs, along with $s(\mathbf{r})$ as the density
of all defects $\Theta$, we have the condition, 
\begin{equation}
s(\mathbf{r})=\bar{s}(\mathbf{r})-a\nabla\cdot\mathbf{D}(\mathbf{r})+a^{2}\sum_{\alpha,\beta}\partial_{\alpha}\partial_{\beta}Q_{\alpha\beta}(\mathbf{r}).\label{eq:SDQ}
\end{equation}

With the concrete configuration of defects $\Theta$ in Eq.~\eqref{rank-1},
the corresponding velocity field $v_{s,\alpha\beta}(\mathbf{r})$
can be evaluated as 
\begin{align}
v_{s,\alpha\beta}(\mathbf{r})= & -\sum_{i}\ell_{i}\delta_{\alpha\beta}\left(-1-\ln\frac{\vert\mathbf{r}-\mathbf{r}_{i}\vert}{a}\right)\nonumber \\
= & -\int d^{2}\mathbf{r}^{\prime}\delta_{\alpha\beta}\left(-1-\ln\frac{\vert\mathbf{r}-\mathbf{r}^{\prime}\vert}{a}\right)s(\mathbf{r}^{\prime})\nonumber \\
= & \int d^{2}\mathbf{q}\delta_{\alpha\beta}\frac{s(\mathbf{q})}{q^{2}}e^{i\mathbf{q}\cdot\mathbf{r}}
\end{align}
which simply indicates a relation between velocity fields $v_{s,\alpha\beta}$
and the defect density $s(\mathbf{q})$: 
 $v_{s,\alpha\beta}(\mathbf{q})=\delta_{\alpha\beta}\frac{s(\mathbf{q})}{q^{2}}
 $.  Therefore, we can alternatively characterize the defect system by
a unified Hamiltonian, 
\begin{align}
H & =\frac{\rho_{0}}{2}\int d^{2}\mathbf{q}\frac{2}{q^{4}}s\left(\mathbf{q}\right)s\left(-\mathbf{q}\right)\,,\label{eq:generalHam}
\end{align}
where $s(\mathbf{q})$ is a density of total defects $\Theta$. From
Eq.~(\ref{eq:generalHam}), one can also recover the defect interaction.
For example, when $s(\mathbf{r})=-a\nabla\cdot\mathbf{D}(\mathbf{r})$,
we have the Hamiltonian (\ref{eq:dbsHam}) for interacting dipole
BSs by adding the core energy term.

\subsection{ Renormalization group analysis of   the first KT transition}\label{section_derivation_of_beta_functions}


To describe how the topological defects  reduce the superfluid stiffness, we have to integrate out the topological defects, which is represented in terms of velocity field $v_{s,\alpha\beta}$. The resulting free energy is 

\begin{equation}
F(\tilde{v}_{\alpha\beta})=-T\ln\int Dv_{s,\alpha\beta}\exp\left(-\beta H\right),\label{eq:free energy}
\end{equation}
where 
\begin{equation}
H=\frac{\rho_{0}}{2}\int d^{2}\mathbf{r}\sum_{\alpha\beta}\left(\tilde{v}_{\alpha\beta}^{2}+2\tilde{v}_{\alpha\beta}v_{s,\alpha\beta}+v_{s,\alpha\beta}^{2}\right),\label{Hamiltonian: velocity-1}
\end{equation}
and $\tilde{v}_{\alpha\beta}$ is the velocity field in the absence
of defects.
The renormalization effect can be measured by the change of a free
energy in response to a small constant velocity shift to the smooth
component 
$\tilde{v}_{\alpha\beta}\rightarrow\tilde{v}_{\alpha\beta}+u_{\alpha\beta}$,
\begin{equation}
F(\tilde{v}_{\alpha\beta}+u_{\alpha\beta})-F(\tilde{v}_{\alpha\beta})=\frac{1}{2}\Omega\rho_{R}\sum_{\alpha\beta}u_{\alpha\beta}^{2}+\mathcal{O}(u_{\alpha\beta}^{2}),\label{free energy}
\end{equation}
where $\rho_{R}$ is the renormalized superfluid stiffness 
\begin{equation}
\rho_{R}=\frac{1}{\Omega}\frac{\delta^{2}F(\tilde{v}_{\alpha\beta}+u_{\alpha\beta})}{\delta u_{\alpha\beta}\delta u_{\alpha\beta}}\Bigg|_{u=0},\label{stiffness}
\end{equation}
and $\Omega$ is the system area. In the dilute limit, after averaging
over defect configuration, we can extract $\rho_{R}$, 
\begin{align}
\rho_{R} & =\rho_{0}-\beta\rho_{0}^{2}\int d^{2}\mathbf{r}\sum_{\alpha\beta}\left\langle v_{s,\alpha\beta}\left(\mathbf{r}\right)v_{s,\alpha\beta}\left(0\right)\right\rangle \nonumber \\
 & =\rho_{0}-\beta\rho_{0}^{2}\lim_{q\rightarrow0}\frac{2}{q^{4}}\left\langle s(\mathbf{q})s(-\mathbf{q})\right\rangle ,\label{stiffness: correlator}
\end{align}
where $\left\langle \boldsymbol{\cdot}\right\rangle $ averages over
all defect configurations and the relation ``$v_{s,\alpha\beta}(\mathbf{q})=\delta_{\alpha\beta}\frac{s(\mathbf{q})}{q^{2}}$''  is adopted.
The formula in Eq.~(\ref{stiffness: correlator}) is our
starting point to derive the RG equation of KT transition. 
The remaining
task is to compute the correlator $\left\langle s(\mathbf{q})s(-\mathbf{q})\right\rangle $.

Topological defects $\Theta$ are confined into quadrupoles in the
low temperature. As the temperature increases, the quadrupole BSs begin to disassociate into dipole BSs at the first
transition and then dipole BSs fully split into freely movable
defects $\Theta$ at the second transition. 
Physically, the creation of an isolated defect or a dipole BS requires
positive chemical potential, thus, we consider the core energy $\mathcal{E}_{s}$
for an isolated defect and the core energy $\mathcal{E}_{d}$ for
a dipole BS. As the formation of the dipole BS will lower the energy,
we have to require $\mathcal{E}_{d}<2\mathcal{E}_{{s}}$ in general.

Around the first transition, dipole BSs can be thermally
activated to reduce the superfluid stiffness. Thus, a system with only dipole BSs is considered here and we need
to evaluate the dipole BS contribution to the correlators. In the
grand ensemble $\Xi=\sum_{\left\{ \mathbf{D}\left(\mathbf{r}\right)\right\} }e^{-\beta H_{D}}$ and 
 Hamiltonian $H_{D}$ is given in Eq. (\ref{eq:dbsHam}). In
real space, 
\begin{align}
H_{D}= & \frac{1}{2}\rho_{0}\int_{\vert\mathbf{r}-\mathbf{r}^{\prime}\vert>a}d^{2}\mathbf{r}d^{2}\mathbf{r}^{\prime}\sum_{\alpha\beta}D_{\alpha}(\mathbf{r})D_{\beta}(\mathbf{r}^{\prime})\nonumber \\
 & \cdot\left(-2\pi\delta_{\alpha\beta}\log\frac{\vert\mathbf{r}-\mathbf{r}^{\prime}\vert}{a}-2\pi\frac{\left(r_{\alpha}-r_{\beta}\right)(r_{\alpha}'-r_{\beta}')}{\vert\mathbf{r}-\mathbf{r}^{\prime}\vert^{2}}\right)\nonumber \\
 & +\mathcal{E}_{d}\ensuremath{\int d^{2}\mathbf{r}}\mathbf{D}^{2}(\mathbf{r}).\label{eq:dipoleHam}
\end{align}
The system should meet the dipole charge
neutral condition $\sum_{i}\mathbf{p}_{i}=0$. In the low energy limit,
since the core energy is large as compared with the temperature, and
the dipole BSs are dilute, we can calculate physical quantity
in a power series of the fugacity $y_{d}=e^{-\beta\mathcal{E}_{d}}$.
With the dipole charge neutral condition and the relation between
$s(\mathbf{q})$ and $\mathbf{D}(\mathbf{q})$, we can express the
defect correlator in terms of dipole BS correlators,
\begin{align}
 & \lim_{q\rightarrow0}\frac{\langle s(\mathbf{q})s(-\mathbf{q})\rangle}{q^{4}}=-\frac{\pi}{2}\sum_{i\neq j}\int_{r_{ij}>a}\frac{dr_{ij}}{a^{2}}\langle\mathbf{p}_{i}\cdot\mathbf{p}_{j}r_{ij}^{3}\rangle,
\end{align}
where $\mathbf{r}_{ij}=\mathbf{r}_{i}-\mathbf{r}_{j}$. In the dilute
limit, we can just consider a pair of dipole BSs. Regarding
the energy in Eq.~(\ref{eq:dipoleHam}), there are five cases in the
leading order, $\mathbf{p}_{i}=(1,0)$,$\mathbf{p}_{j}=(-1,0)$ or $\mathbf{p}_{i}=(-1,0)$,$\mathbf{p}_{j}=(1,0)$
or $\mathbf{p}_{i}=(0,1)$, $\mathbf{p}_{j}=(0,-1)$ or $\mathbf{p}_{i}=(0,-1)$,
$\mathbf{\mathbf{p}}_{j}=(0,1)$ or $\mathbf{p}_{i}=\mathbf{p}_{j}=(0,0)$. With a fixed separation
$\mathbf{r}_{ij}$, the first four cases have equal contributions
to grand ensemble. We hence obtain 
 $\langle\mathbf{p}_{i}\cdot\mathbf{p}_{j}r_{ij}^{3}\rangle=-4y_{d}^{2}\left(\frac{r_{ij}}{a}\right)^{3-2\pi\beta\rho_{0}},
 $  
which yields 
\begin{align}
\lim_{q\rightarrow0}\frac{\langle s(\mathbf{q})s(-\mathbf{q})\rangle}{q^{4}} & =2\pi y^{2}\int_{r>a}\frac{dr}{a}(\frac{r}{a})^{3-2\pi\beta\rho_{0}}+\mathcal{O}(y_{d}^{4}).\label{eq:<ss>}
\end{align}
We are ready to send Eq.~(\ref{eq:<ss>}) back to Eq.~(\ref{stiffness: correlator}).
After keeping the lowest order in $\rho_{R}^{-1}$, we have 
\begin{equation}
\rho_{R}^{-1}=\rho_{0}^{-1}-2\pi\int_{r>a}\frac{dr}{a}(\frac{r}{a})^{3-2\pi\beta\rho_{0}}+\mathcal{O}(y_{d}^{4}).
\end{equation}
The RG equation arises when we enforce the standard regularization
scheme by dividing the integral into two parts, $\int_{r>a}=\int_{a}^{ae^{\delta l}}+\int_{ae^{\delta l}}^{\infty}$
which leads to two equations, 
\begin{equation}
\rho_{R}^{-1}=\rho^{-1}\left(\delta l\right)+2\pi\beta y_{d}^{2}\int_{ae^{\delta l}}^{+\infty}\frac{d\xi}{a}\left(\frac{\xi}{a}\right)^{3-2\pi\beta\rho_{0}}+\mathcal{O}\left(y_{d}^{4}\right)\label{stiffness: large xi-1}
\end{equation}
and 
\begin{align}
\rho^{-1}\left(\delta l\right) & =\rho_{0}^{-1}+2\pi\beta y_{d}^{2}\int_{a}^{ae^{\delta l}}\frac{d\xi}{a}\left(\frac{\xi}{a}\right)^{3-2\pi\beta\rho_{0}}+\mathcal{O}\left(y_{d}^{4}\right).\label{stiffness: small xi-1}
\end{align}
Substituting Eq.~(\ref{stiffness: small xi-1}) into Eq.~(\ref{stiffness: large xi-1})
yields the renormalized fugacity, $y(\delta l)=ye^{(2-\beta\pi\rho_{0})\delta l}$. Therefore,
we reach the RG equations (\ref{eq:rhorg}) and (\ref{eq:yeq}) for the first transition at $T_{c1}$.   
Above $T_{c1}$, the superfluid stiffness gets renormalized into $\rho_R\rightarrow 0$,
indicating that the superfluid phase is broken down. We then goes
to a phase with the dipole BSs as freely mobile charge carriers. 
In deriving the RG equations, we have introduced a lattice regularization
on which the dipole charges can only take integer values. Also, one
can repeat the above procedures in a continuum space, in which $\mathbf{p}_{i}\cdot\mathbf{p}_{j}=\ell_{i}\ell_{j}\boldsymbol{\xi}_{i}\cdot\boldsymbol{\xi}_{j}$
includes a part from the relative angular $\boldsymbol{\xi}_{i}\cdot\boldsymbol{\xi}_{j}=\xi_{i}\xi_{j}\cos(\theta_{ij})$. It
will alter the coefficient before $y^{2}$,
but the physics will not be changed.

\section{Phase-III: defects are deconfined}\label{section_phase3}

\subsection{General properties of Phase-III}\label{section_phase3_suppl}
When the temperature is close to the topological phase transition (named as ``the second KT transition'') from Phase-II to Phase-III,
dipole BSs form a plasma due to   sufficiently high density and equivalently we 
have a condensate of dipole BSs. Therefore,
the interaction between two defects $\Theta$, i.e., $\mathcal{I}$ in Eq.~(\ref{eq:screenedU11}) gets strongly screened.
This dynamical renormalization can be quantitatively treated via the standard  Debye-H{\"u}ckel approximation by approximating 
the dipole BSs as a continuous independent field. 
At  low energies and small $\mathbf q$, we keep the leading orders, yielding an effective Hamiltonian for defects: 
\begin{equation}
 {H}_{\text{eff}}=\int d^{2}\mathbf{q}\left[\frac{\mathcal{E}_{d}}{q^{2}}+(\mathcal{E}_{s}-\rho_{0}\lambda_{d}^{4})\right]s(\mathbf{q})s(-\mathbf{q}),\label{eq:dybehukel}
\end{equation}
where $s(\mathbf{q})$ is the Fourier transformation of a defect density
$s(\mathbf{r})\equiv\sum_{i}\ell_{i}\delta(\mathbf{r}-\mathbf{r}_{i})$
and $\lambda_{d}=\sqrt{\mathcal{E}_{d}/\rho_{0}}$ denotes the Debye-H{\"u}ckel
screening length. 
The detailed derivation of this Hamiltonian is shown
in Sec.~\ref{subsec:Debye}.
Physically, the dipole BS plasma brings two significant 
effects: 1) to renormalize
the interaction from Eq.~(\ref{eq:screenedU11}) to $\ln\frac{\vert\mathbf{r}_{i}-\mathbf{r}_{j}\vert}{a}$; 2) to reduce the core energy to $\mathcal{E}_{s}-\rho_{0}\lambda_{d}^{4}$. Based on $H_{\text{eff}}$, we can obtain the phase diagram in Fig.~\ref{Fig:phasediagram} via the following analysis of  three aspects:
\begin{enumerate}
\item  When the effective core energy $\mathcal{E}_{s}-\rho_{0}\lambda_{d}^{4}>0$ along with $T_{c2}>T_{c1}$ gives $(2\mathcal{E}_{d})^{-1}<\rho^{-1}_{0}<\mathcal{E}_{s}/\mathcal{E}_{d}^{2}$, i.e., the domain $(A,B)$ in Fig.~\ref{Fig:phasediagram}, 
the system in Eq.~\eqref{eq:dybehukel} reduces to a 2D 
Coulomb gas, which has a KT transition at 
\begin{equation}
  T_{c2}=\pi\mathcal{E}_{d}\,.\label{eq_tc2}
\end{equation} 
The second KT transition $T_{c2}$ merely depends on the core energy of dipole BSs, which 
corresponds to the horizontal line separating Phase II and  III in Fig.~\ref{Fig:phasediagram}.

\item Instability occurs in case of a negative reduced core energy $\mathcal{E}_{s}-\rho_{0}\lambda_{d}^{4}<0$ (i.e., $\rho^{-1}_0>\mathcal{E}_{s}/\mathcal{E}^2_d$) or $T_{c2}<T_{c1}$ (i.e., $\rho^{-1}_{0}<(2\mathcal{E}_{d})^{-1}$). In the former case, instead,  the screening
effect favors a finite density of defects $\Theta$, 
 $\langle s(\mathbf{q})\rangle\text{\ensuremath{\neq0}}$, when  higher-order term is involved in 
Eq.~(\ref{eq:dybehukel}), with a vertical line as a boundary between Phase II and Phase III. In the latter case, Phase II is metastable and the two transitions merge together.
 In both cases, the intermediate Phase-II between Phase-I and Phase-II
does not exist and we end up with  a direct transition  from Phase-I to Phase-III, where  quadrupole BSs are directly dissolved into free   defects.

  \item At the high temperature limit in Phase-III where defects $\Theta$
reach a sufficiently high density,
a plasma phase of isolated defects forms. The effective interaction
of defects in this phase will be renormalized by the susceptibility $\chi_{ss}\left(\mathbf{q}\right)$
induced by defect plasma. The effective interaction is 
\begin{align}
\frac{2\rho_{0}}{q^{4}}-\rho_{0}\chi_{ss}\left(\mathbf{q}\right) & =\frac{2\rho_{0}}{q^{4}+\lambda_{s}^{-4}},\label{eff inteteraction}
\end{align}
where $\lambda_{s}$ is the temperature-dependent Debye-H\"uckel screening
length of defects $\Theta$ and we can find that the screening effect
leads to a short ranged interaction. Thus, in Phase-III, defects
are called Calypso in Introduction. We leave the concrete derivation of the effective interaction in Sec.~\ref{subsec:defect plasma}. 
 \end{enumerate}
\subsection{  Debye-H\"uckel approximation and   the second KT transition  \label{subsec:Debye} }

Around the second transition point, the density of quadrupole BSs vanishes, $ {Q}_{\alpha\beta}=0$. 
At the stage, the Hamiltonian only involves isolated defect density
$\bar{s}(\mathbf{r})$ and dipole BS density $\mathbf{D}(\mathbf{r})$,
\begin{equation}
H=\frac{\rho_{0}}{2}\int d^{2}q\frac{2}{q^{4}}s(\mathbf{q})s(-\mathbf{q})+\int d^{2}r\left[\mathcal{E}_{s}\bar{s}^{2}(\mathbf{r})+\mathcal{E}_{d}\mathbf{D}^{2}(\mathbf{r})\right],\label{eq:Hamsd}
\end{equation}
where the total defect density $s(\mathbf{r})=\bar{s}(\mathbf{r})-\nabla\cdot\mathbf{D}(\mathbf{r})$.
The dipole BSs have sufficiently high density to form a plasma, which
allows us to apply the Debye-Hückel approximation and to take the density $\mathbf{D}(\mathbf{r})$ as a continuous
function.
The first terms in Eq. (\ref{eq:Hamsd}) contains the coupling terms
between isolated defects $\bar{s}(\mathbf{q})$ and dipole BSs $\mathbf{D}(\mathbf{q})$, i.e. 
\begin{align}
\frac{1}{q^{4}}s(\mathbf{q})s(-\mathbf{q})= & \frac{1}{q^{4}}\bar{s}(\mathbf{q})\bar{s}(-\mathbf{q})+2\sum_{\alpha}\frac{iq_{\alpha}}{q^{4}}\bar{s}(\mathbf{q})D_{\alpha}(-\mathbf{q})\nonumber \\
 & +\sum_{\alpha\beta}\frac{q_{\alpha}q_{\beta}}{q^{4}}D_{\alpha}(\mathbf{q})D_{\beta}(-\mathbf{q})\label{eq:coupling}\,.
\end{align}
The plasma of dipole BSs can screen the interaction between
defects. We first evaluate the correlator of dipole bound
state.
Notice
that the partition function for the dipole BSs is  $\Xi_{d}=\int\mathcal{D}\mathbf{D}\exp(-\beta H_{d})\,,
 $ where 
\begin{equation}
H_{d}=\int d^{2}q\sum_{\alpha\beta}\left(\rho_{0}\frac{q_{\alpha}q_{\beta}}{q^{4}}+\delta_{\alpha\beta}\mathcal{E}_{d}\right)D_{\alpha}(\mathbf{q})D_{\beta}(-\mathbf{q})
\end{equation}
Here, $\mathbf{D}(\mathbf{q})$ is a continuous function by taking
the Debye-H\"uckle approximation. Thus, the correlator can be calculated
by Gaussian integral, 
\begin{align}
 & \left\langle D_{\alpha}\left(\mathbf{q}\right)D_{\beta}\left(-\mathbf{q}\right)\right\rangle \nonumber \\
= & \frac{1}{\mathcal{E}_{d}}\left[\left(\delta_{\alpha\beta}-\frac{q_{\alpha}q_{\beta}}{q^{2}}\right)+\frac{q_{\alpha}q_{\beta}}{q^{2}}\frac{q^{2}}{\lambda_{d}^{-2}+q^{2}}\right]\,,\label{eq:ddcor}
\end{align}
where $\lambda_d=\sqrt{\mathcal{E}_{d}/\rho_{0}}$ is the Debye-H\"uckel
screening length of dipole BSs. 

 The effective interaction in the dipole
BS plasma is renormalized by correlator in Eq.~(\ref{eq:ddcor}),
\begin{equation}
\frac{\rho_{0}}{q^{4}}-\sum_{\alpha\beta}\frac{\rho_{0}^{2}q_{\alpha}q_{\beta}}{q^{8}}\left\langle D_{\alpha}\left(\mathbf{q}\right)D_{\beta}\left(-\mathbf{q}\right)\right\rangle =\frac{\rho_{0}}{q^{2}}\frac{1}{\lambda_{d}^{-2}+q^{2}}
\end{equation}
where the factor $q_{\alpha}q_{\beta}/q^{8}$ is from the coupling
term in Eq. (\ref{eq:coupling}). With the core energy $\mathcal{E}_{s}$
for isolated defects, the Hamiltonian for defects in dipole BS plasma is 
\begin{align}
H & =\int d^{2}q\left(\frac{\rho_{0}}{q^{2}}\frac{1}{\lambda_{d}^{-2}+q^{2}}+\mathcal{E}_{s}\right)\bar{s}(\mathbf{q})\bar{s}(-\mathbf{q})\label{eq:Hscreen}
\end{align}
Making the expansion in the low energy will result in the effective
Hamiltonian in Eq.~(\ref{eq:dybehukel}).

\subsection{The defect plasma and and the effective interaction\label{subsec:defect plasma}}
 
At high temperature $T\gg T_{c2}$ region, both dipole and quadrupole
BSs disassociate into free isolated defects. This is a plasma phase of defects, and then the interactions between defects 
will get screened. In this case, the Hamiltonian 
just involves density of
isolated defects, which reads 
\begin{equation}
H=\frac{\rho_{0}}{2}\int d^{2}\mathbf{q}\frac{2}{q^{4}}s\left(\mathbf{q}\right)s\left(-\mathbf{q}\right)+\mathcal{E}_{\text{s}}\int d^{2}\mathbf{r}s^{2}\left(\mathbf{r}\right).\label{Hamiltonian: isolated defect-3}
\end{equation}
We regularize the system with the constant by a lattice with a lattice
constant $a$. Then a plasma phase is referred to as one defect per
site, 
 $
\sum_{\ell\neq0}n_{\ell}\left(\mathbf{r}\right)=1$, 
where $n_{\ell}\left(\mathbf{r}\right)$ is the number of a defect
with charge $\ell$ at site $\mathbf{r}$, $n_{\ell}\left(\mathbf{r}\right)=\sum_{i}\delta_{\ell,\ell_{i}}\delta\left(\mathbf{r}-\mathbf{r}_{i}\right)$
with a relation $\bar{s}\left(\mathbf{r}\right)=\sum_{\ell}\ell n_{\ell}\left(\mathbf{r}\right)$.
Then, we reformulate the Hamiltonian in (\ref{Hamiltonian: isolated defect-3})
\begin{align}
H=&\frac{\rho_{0}}{2}\int d^{2}\mathbf{r}d^{2}\mathbf{r}'\sum_{\ell,\ell'}\ell\ell'n_{\ell}\left(\mathbf{r}\right)U\left(\mathbf{r}-\mathbf{r}'\right)n_{\ell'}\left(\mathbf{r}'\right)\nonumber\\
&+\mathcal{E}_{s}\int d^{2}\mathbf{r}\sum_{\ell}\ell^{2}n_{\ell}\left(\mathbf{r}\right),\label{Hamiltonian: isolated defect, number density}
\end{align}
where $U\left(\mathbf{r}\right)$ is the Fourier transformation of
$2/q^{4}$. $n_{\ell}\left(\mathbf{r}\right)$ is a function of the
 position $\mathbf{r}_{i}$ of the individual defects. In order to
obtain the interaction screened by the plasma, we decompose all the
defects $n_{\ell}\left(\mathbf{r}\right)$ into the continuous plasma
field $\left\langle n_{\ell}\left(\mathbf{r}\right)\right\rangle $
and the additional defects placed in the plasma $\delta n_{\ell}\left(\mathbf{r}\right)$,
 $ n_{\ell}\left(\mathbf{r}\right)=\left\langle n_{\ell}\left(\mathbf{r}\right)\right\rangle +\delta n_{\ell}\left(\mathbf{r}\right)$.    
Keeping the terms containing $\left\langle n_{\ell}\left(\mathbf{r}\right)\right\rangle $,
we have the free energy at the mean-field level \cite{chaikin2000principles}
\begin{widetext}
\begin{align}
F= & \frac{\rho_{0}}{2}\int d^{2}\mathbf{r}d^{2}\mathbf{r}'\sum_{\ell,\ell'}\ell\ell'\left\langle n_{\ell}\left(\mathbf{r}\right)\right\rangle U\left(\mathbf{r}-\mathbf{r}'\right)\left\langle n_{\ell'}\left(\mathbf{r}'\right)\right\rangle   -\int d^{2}\mathbf{r}\sum_{\ell}\ell\left\langle n_{\ell}\left(\mathbf{r}\right)\right\rangle \phi^{\text{ext}}\left(\mathbf{r}\right)\nonumber \\
 & +\mathcal{E}_{s}\int d^{2}\mathbf{r}\sum_{\ell}\ell^{2}\left\langle n_{\ell}\left(\mathbf{r}\right)\right\rangle +\frac{1}{\beta}\sum_{\ell}\int d^{2}\mathbf{r}\left\langle n_{\ell}\left(\mathbf{r}\right)\right\rangle \ln\left\langle n_{\ell}\left(\mathbf{r}\right)\right\rangle ,\label{free energy: mean field}
\end{align}
with an external field $\phi^{\text{ext}}\left(\mathbf{r}\right)=\sum_{\ell}\ell\delta n_{\ell}\left(\mathbf{r}'\right)U\left(\mathbf{r}-\mathbf{r}'\right)$.
In the equilibrium, minimize $F$ under the condition  $
\sum_{\ell\neq0}n_{\ell}\left(\mathbf{r}\right)=1$ and we obtain the mean field solution
\begin{equation}
\left\langle n_{\ell}\left(\mathbf{r}\right)\right\rangle =\frac{\exp\left[\beta\ell\phi^{\text{ext}}\left(\mathbf{r}\right)-\beta\ell^{2}\mathcal{E}_{s}-\beta\ell\rho_{0}\int d^{2}\mathbf{r}'\sum_{\ell'\neq0}\ell'\left\langle n_{\ell'}\left(\mathbf{r}'\right)\right\rangle U\left(\mathbf{r}-\mathbf{r}'\right)\right]}{\sum_{\ell\neq0}\exp\left[\beta\ell\phi^{\text{ext}}\left(\mathbf{r}\right)-\beta\ell^{2}\mathcal{E}_{s}-\beta\ell\rho_{0}\int d^{2}\mathbf{r}'\sum_{\ell'\neq0}\ell'\left\langle n_{\ell'}\left(\mathbf{r}'\right)\right\rangle U\left(\mathbf{r}-\mathbf{r}'\right)\right]}.\label{density expression}
\end{equation}
\end{widetext}
The effective interaction for $\delta n_{\ell}(\mathbf{r})$ can be
yielded by the linear response function 
\begin{equation}
\chi_{ss}\left(\mathbf{r},\mathbf{r}'\right)=\sum_{\ell}\ell\frac{\delta\left\langle n_{\ell}\left(\mathbf{r}\right)\right\rangle }{\delta\phi^{\text{ext}}\left(\mathbf{r}'\right)}\Bigg|_{\phi^{\text{ext}}=0}.\label{response function}
\end{equation}

In response to fluctuation field $\phi^{\text{ext}}\left(\mathbf{r}\right)$,
the $\left\langle n_{\ell}\left(\mathbf{r}\right)\right\rangle $
is changed according to Eq.~(\ref{density expression}). We can evaluate
the susceptibility $\chi_{ss}\left(\mathbf{r},\mathbf{r}'\right)$
as 
\begin{align}
\chi_{ss}\left(\mathbf{r},\mathbf{r}'\right)= & \frac{\rho_{0}\mathcal{E}_{s}}{2\lambda_{s}^{4}\mathcal{E}_{c}}\Bigl[\delta\left(\mathbf{r}-\mathbf{r}'\right)\nonumber \\
 & -\rho_{0}\int d^{2}\mathbf{r}''\chi_{ss}\left(\mathbf{r}'',\mathbf{r}'\right)U\left(\mathbf{r}-\mathbf{r}''\right)\Bigr],
\end{align}
and in the momentum space, 
 \begin{align}
\chi_{ss}\left(\mathbf{q}\right)=\left(\frac{2\rho_{0}}{q^{4}}+\frac{2\rho_{0}\lambda_{s}^{4}\mathcal{E}_{c}}{\mathcal{E}_{s}}\right)^{-1}, 
 \end{align}
with a coefficient  $\lambda_{s}=\left(\frac{2\rho_{0}\sum_{\ell\neq0}\beta\mathcal{E}_{c}\ell^{2}e^{-\beta\mathcal{E}_{s}\ell^{2}}}{\mathcal{E}_{s}\sum_{\ell\neq0}e^{-\beta\mathcal{E}_{\text{s}}\ell^{2}}}\right)^{-1/4}. $  
In the high temperature limit, we have $\beta\mathcal{E}_{\text{s}}\rightarrow0$
and $\lambda_{s}=\left(\rho_{0}/\mathcal{E}_{s}\right)^{-1/4}$. 
With
susceptibility $\chi_{ss}\left(\mathbf{q}\right)$ describing the
screening effect, the interaction of defects reduces to a short ranged
interaction in Eq.~(\ref{eff inteteraction}).

\section{Generalization}\label{section_generalization}
The above discussion  
can be generalized to defects of a general higher-rank symmetry group with multi-moment conservation.
As a result, a minimal bound state formed by defects carrying
opposite topological charges to reach neutralization, resembles a local
particle with a finite energy. Consider a system containing $N$ defects
at $\mathbf{r}_{i},i=1,\ldots,N$ with topological charges $\ell_{i},i=1,\ldots,N$.
The total energy can be formulated as 
\begin{align}
\mathscr{E}=\sum_{i,j}\ell_{i}\ell_{j}\int d^{2}\mathbf{q}\,\mathscr{U}_\mathbf{q}e^{i\mathbf{q}\cdot(\mathbf{r}_{i}-\mathbf{r}_{j})},
\end{align}
where $\mathscr{U}_\mathbf{q}$ describes defect interaction and generally
we assume $\mathscr{U}_\mathbf{q}\propto1/q^{2n}$
($n\in\mathbb{Z}$) at low energies, i.e., the IR limit with small
momentum $q$, which leads to energy divergence of a single defect.
We can expose the divergent features: 
\begin{align}\int d^{2}\mathbf{q}\,\mathscr{U}_\mathbf{q}e^{i\mathbf{q}\cdot(\mathbf{r}_{i}-\mathbf{r}_{j})}\propto\int d^{2}\mathbf{q}\frac{P_{2n-2}}{q^{2n}}+\mathrm{finite}
\end{align}
with $P_{n}(\mathbf{r})=\sum_{k=0}^{n}\frac{(i\mathbf{q}\cdot\mathbf{r})^{k}}{k!}$
and finite denoting the remaining non-divergent terms. By inspecting the power-law and logarithmic divergence, we have the BS that can be thermally activated from the ground states with conservation of the first $m$-th moments
\begin{align}Q_{\alpha_{1}\alpha_{2}\cdots\alpha_{n}}\equiv\sum_{i}\ell_{i}r_{i\alpha_{1}}r_{i\alpha_{2}}\cdots r_{i\alpha_{n}}=0,n\leq m. 
\end{align}
We fix our attention on defects carrying
topological charges $\pm1$, and at low energies, the minimal
BS with the first $2n$th moments vanishing, contain $2^{n}$
defects, with half $\ell=+1$ and half $\ell=-1$ and the
spatial alignment of these vortices is strongly constrained.

Accordingly, we can establish  a series of KT topological
transitions that successively occur at different temperatures due to proliferation of defects and defect BSs. 
The effect of hierarchical proliferation can be theoretically characterized by means of RG flow equations and the Debye-H\"uckel approximation. More interestingly, a screening effect from sufficiently high density of defect BSs can lead to instability and collapse of the intermediate temperature phases, which further enriches the phase diagrams. In particular, the first
KT point is determined by the stiffness, while the other ones are
related to the BS core energies.

\section{Summary and Outlook}\label{section_summary}

 As an important ingredient of the series of works on fractonic superfluids, in this paper, we have  systematically investigated the finite-temperature phase structure of fractonic superfluids, where the conventional $\mathbb{U}(1)$ symmetry is replaced by higher rank symmetry and the conventional superfluid vortex is replaced by higher rank symmetry defect. A hierarchical proliferation of topological objects  has been identified, resulting in a finite-temperature phase diagram shown in Fig.~\ref{Fig:phasediagram}. 
The model in Eq.~\eqref{eq:Ham} can be promisingly realizable in the cold atomic gas subjected to an optical lattice by tuning a two-particle state \cite{ExperimentFisher2005}.  Eq.~(\ref{eq_tc2}) indicates that $T_{c2}$ is determined by $\mathcal{E}_d$, i.e., the core energy of dipole-like BSs of defects.   Naively, a core energy manifests as the increase in free energy due to destruction of the superfluidity, i.e., $\mathcal E_d\sim a^2 f_\mathrm{cond}$ with $f_\mathrm{cond}$ being the superfluid energy density. Along with the Josephson scaling relation $\rho(T)\sim T_c\xi$ and $f_\mathrm{cond}\sim T_c\xi^{-2}$, we can estimate the core energy $\mathcal E_d\sim \pi\rho_0/2$ at the optimal valu{}e of $a$ that minimizes the total energy $\mathscr{E}^d_{\mathrm{self}}+\mathcal E_d$ of a single dipole BS with a unit charge.   Simultaneously, a numerical simulation on a proper lattice regularized model is desired to benchmark the phase diagram in Fig.~\ref{Fig:phasediagram} by incorporating the defect core physics. One may also decorate these defects \cite{Xie2014decorated} and then
defect condensation \cite{yp18prl,PhysRevB.93.115136,bti2,YeGu2015}
may generate new classes of Symmetry Protected Topological (SPT) phases
beyond group cohomology~\cite{Chen_science}. The non-equilibrium
properties are also of fundamental importance, since the high rank
symmetry enforces dynamics with constraint kinetics. Example is the
Kibble-Zurek mechanism of the forming of high-rank defects that is
triggered through a continuous phase transition at finite rate. Recently,
HRS has attracted much attention from fields other than condensed
matter physics, see, e.g., Refs.~\cite{PhysRevD.104.105001,Bidussi:2021nmp,Jain:2021ibh,Angus:2021jvm,Grosvenor:2021hkn}.
It is   interesting to study effect of defects.  The corresponding response theory may involve exotic geometric (gravitational) response, leading to, potentially, exotic topological terms, e.g., mixture of Riemann-Cartan quantities and fibre bundle of internal gauge fields, and twisted topological terms \cite{PhysRevB.99.205120,bti6,PhysRevResearch.3.023132,PhysRevLett.112.141602,Ye:2017aa,PhysRevB.93.115136,yp18prl,2016arXiv161209298P,Zhang:2022rbg}.

\acknowledgements 
 The first two authors
(J.K.Y. and S.A.C.) contributed equally to this work. The authors
acknowledge helpful discussions with T.K. Ng, K.T. Law and Y.B. Yang.
This work was supported by Guangdong Basic and Applied Basic Research
Foundation under Grant No.~2020B1515120100, NSFC Grant (No.~12074438).  The work reported here was performed on resources provided in part by the Guangdong Provincial Key Laboratory of Magnetoelectric Physics and Devices (LaMPad).


%


\end{document}